\documentclass[twocolumn]{revtex4}               %start writting 27 June 12%
\usepackage{amsmath,amssymb}
\usepackage{latexsym}
\usepackage{epsfig}

\begin{document}
\title{Tsallis holographic dark energy in the brane cosmology}
\author{S. Ghaffari$^{1}$\footnote{sh.ghaffari@riaam.ac.ir}, H. Moradpour$^1$\footnote{h.moradpour@riaam.ac.ir}, J. P. Morais Gra\c
ca$^{2}$\footnote{jpmorais@gmail.com}, Valdir B.
Bezerra$^{2}$\footnote{valdir@fisica.ufpb.br}, I. P.
Lobo$^{2}$\footnote{iarley\_lobo@fisica.ufpb.br}}
\address{$^1$ Research Institute for Astronomy and Astrophysics of Maragha (RIAAM), P. O. Box 55134-441, Maragha, Iran\\
$^{2}$ Departamento de F\'{i}sica, Univeidade Federal da
Para\'{i}ba, Caixa Postal 5008, CEP 58051-970, Jo\~{a}o Pessoa, PB,
Brazil}

\begin{abstract}
We study some cosmological features of Tsallis
holographic dark energy (THDE) in Cyclic, DGP and RS II braneworlds.
In our setup, a flat FRW universe is considered filled by a
pressureless source and THDE with the Hubble radius as the IR
cutoff, while there is no interaction between them. Our result shows
that although suitable behavior can be obtained for the system
parameters such as the deceleration parameter, the models are not
always stable during the cosmic evolution at the classical level.
\end{abstract}
 \maketitle

\section{Introduction}

Due to the weakness of general relativity to describe the current
accelerated universe \cite{Riess,roos}, physicists try to eliminate
this difficulty by $i$) introducing amazing energy sources, called
dark energy, $ii$) modifying the general relativity theory or even a
combination of these. Braneworld scenario is an interesting
approach to modify the Einstein theory, and Dvali-Gabadadze-Porrati
(DGP) braneworld, the second model of Randall and Sundrum (RS II)
and the Cyclic model of Steinhardt and Turok are three pioneering
models in this regard \cite{dgp,18,Steinhardt}. There is also
another Cyclic model motivated by both the braneworld and loop
quantum cosmology scenarios \cite{Ashtekar,bc,Brown. Freese,Baum
Frampton,X.Zhang}. The basic idea behind the braneworld hypothesis
is that our universe is a brane embedded in a higher dimensional
bulk, while only gravity can penetrate the bulk, and as well as the
energy-momentum distribution, other forces are limited to the brane
\cite{dgp,18}.

In the DGP braneworld model the $4$-dimensional FRW universe is
embedded in a $5$D Minkowski bulk. DGP braneworld has two branches
of solutions corresponding to $\epsilon=+1$ and $\epsilon=-1$.
Although, the first case provides a self-accelerating solution for
the current universe, it suffers from the ghost instability problem
\cite{Koyama}. The normal branch of $\epsilon=-1$ requires dark
energy to describe the accelerated universe. On the other hand, the
idea that our universe may consist of an infinite cycle of expansions
and contractions leads to an interesting model for the universe
called the cyclic universe \cite{Tolman}. A new version of this
model has been proposed \cite{Ashtekar,bc} which suffers from two
main problems \cite{Brown. Freese,Baum Frampton,X.Zhang}. These
problems, including the black hole and entropy problems \cite{Brown.
Freese,Baum Frampton,X.Zhang}, are solved by considering the phantom
dark energy (PDE) \cite{Brown. Freese,Baum Frampton}.

Holographic principle permits us to establish an upper bound for the
energy density of quantum fields in vacuum \cite{Cohen}. Using the
Bekenstein entropy and this principle, a model for dark energy has
been proposed called holographic dark energy (HDE) and suffers from
the stability problem \cite{HDE1,HDE2,HDE3,HDE4,HDE5,stab}. This
idea has been employed in order to model dark energy by the energy
density of quantum fields in vacuum, in the DGP, RSII and cyclic
universes \cite{DGP,Ghaffari,RS,Saridakis,Cyclic1,Cyclic2}.

Since gravity is a long-range interaction, it may satisfy the
non-extensive probability distributions \cite{non3}. This view leads
to interesting results in gravitational and cosmological setups
\cite{non1,non2,non30,non4,non5,non6,non7,non8,non9,non10,non11}.
Recently, using the Tsallis generalized entropy \cite{non3} and
holographic hypothesis, a new holographic model for dark energy has
been introduced, in which the Hubble radius plays the role of the IR
cutoff, as \cite{Tavayef}

\begin{equation}\label{rhol}
\rho_D=BH^{4-2\delta},
\end{equation}

\noindent where $H=\frac{\dot{a}}{a}$ is the Hubble parameter. The
cosmological features of this dark energy model in various
cosmological setups can be found in Refs.
\cite{THDE1,THDE2,THDE3,THDE4}.

Here, we are interested in studying some cosmological consequences
of employing Eq.~(\ref{rhol}) in the DGP \cite{dgp}, RS II \cite{18}
and Cyclic \cite{Ashtekar,bc} models. Sine the WMAP data indicates a
flat FRW universe, we consider a flat FRW universe, in which there
is no mutual interaction between the cosmos sectors. In order to
achieve this goal, we study some cosmological features of THDE in
Cyclic model in the next section. Secs.~($3$) and~($4$) include its cosmological consequences in the DGP and RS II
braneworlds, respectively. The classical stability of the models are
also studied in the $5$th section. The last section is devoted to a
summary.

%%%%%%%%%%%%%%%%%%%%%%%%%%%%%%%%%%%%%%%%%%%%%%%%%%%%%%%%
\section{THDE in Cyclic Universe}

Effective field theory of loop quantum cosmology modifies the
Friedmann equation as \cite{Ashtekar}

\begin{equation}\label{Friedeq4}
H^2=\frac{\rho}{3m_p^2}(1-\frac{\rho}{\rho_c}),
\end{equation}

\noindent where $\rho$ is the total energy density of the fluid
filling the cosmos, and $\rho_c$ denotes the critical density
constrained by quantum gravity and different from the usual critical
density ($\rho_{cr}=3m_p^2H^2$). This modified Friedmann equation
can also be obtained in the framework of braneworld scenario
\cite{bc,X.Zhang}. In our model, the cosmos includes also dark
matter (DM) and DE, which do not interact mutually, and hence, the total energy-momentum conservation law is
decomposed as

\begin{eqnarray}
&&\dot{\rho}_D+3H(1+\omega_D)\rho_D=0,\label{ConserveDE}\\
&&\dot{\rho}_m+3H\rho_m=0\rightarrow\rho_m=\rho_0(1+z)^3,\label{ConserveCDM}
\end{eqnarray}

\noindent where $\rho_0$ is an integral constant, and we used the
$1+z=\frac{1}{a}$ relation between the redshift $z$ and the scale
factor $a$ while its current time values has been normalized to one.
$\rho_m$ and $\rho_D$ also denote the energy density of DM and DE, respectively, and $\omega_D$ is the equation of state
(EoS) parameter of dark energy. We define the dimensionless density
parameters as

\begin{eqnarray}\label{Omega}
\Omega_m=\frac{\rho_m}{\rho_{cr}}=\frac{\rho_m}{3m_p^2H^2} ~~~
\Omega_D=\frac{\rho_D}{\rho_{cr}}=\frac{\rho_D}{3m_p^2H^2},
\end{eqnarray}

\noindent and insert them in Eq.~(\ref{Friedeq4}) to obtain

\begin{eqnarray}\label{a}
\Omega_D=(1-\Omega_m)+\frac{\Omega_D+\Omega_m}{\frac{\rho_c}{\rho_{cr}}-(\Omega_D+\Omega_m)}.
\end{eqnarray}

\noindent The use of Eqs.~(\ref{ConserveCDM}) and~(\ref{a}) leads to

\begin{eqnarray}\label{a1}
\Omega_D(z\rightarrow-1)\approx1+\frac{\Omega_D}{\frac{\rho_c}{\rho_{cr}}-\Omega_D}.
\end{eqnarray}

\noindent at the $z\rightarrow-1$ limit. This equation clearly
indicates that, at this limit, we have $\Omega_D>1$, if
$\rho_c>\rho_D$ (see Ref. \cite{Cyclic1} for more details). Now,
combining Eq.~(\ref{Omega}) with Eq.~(\ref{rhol}), we find

\begin{equation}\label{Omegac1}
\Omega_D=\frac{B}{3m_p^2}H^{2-2\delta},
\end{equation}

\noindent for the DE dimensionless density parameter. Now, defining
$u=\frac{\Omega_m}{\Omega_D}$, using the time derivative of
Eq.~(\ref{Friedeq4}), and combining the results with
Eqs.~(\ref{Omega}) and~(\ref{Friedeq4}), we arrive at

\begin{equation}\label{Hdot1}
\frac{\dot{H}}{H^2}=\frac{-3u\Big(2-\Omega_D(1+u)\Big)}{2(\delta-2)(2-\Omega_D(1+u))+2(u+1)},
\end{equation}

\noindent which can finally be used to write

\begin{equation}\label{Omegac2}
\Omega^{\prime}_D=\frac{d\Omega_D}{d\ln
a}=\frac{\dot{\Omega}_D}{H}=\frac{-3u(1-\delta)\Omega_D\Big(2-\Omega_D(1+u)\Big)}{(\delta-2)(2-\Omega_D(1+u))+u+1},
\end{equation}

\noindent where dot denotes derivative with respect to time.
Calculations of the EoS and parameter of THDE and deceleration
parameter also lead to

\begin{equation}\label{w1}
\omega_D=-1+\frac{u(2-\delta)\Big(2-\Omega_D(1+u)\Big)}{(\delta-2)(2-\Omega_D(1+u))+u+1}.
\end{equation}

\noindent and

\begin{equation}\label{q1}
q=-1-\frac{\dot{H}}{H^2}=-1+\frac{3u\Big(2-\Omega_D(1+u)\Big)}{2(\delta-2)(2-\Omega_D(1+u))+2(u+1)},
\end{equation}

\noindent respectively.
\begin{figure}[htp]
    \begin{center}
    \includegraphics[width=8cm]{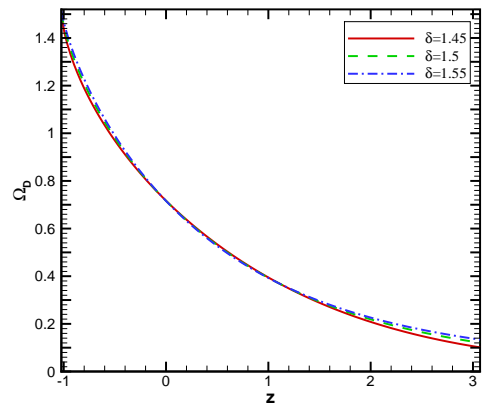}
    \caption{$\Omega_D$ versus $z$ for $\Omega_{D0}=0\cdot73$, $u_0=0\cdot3$, and some values of $\delta$.}\label{Omega1}
    \end{center}
\end{figure}
\begin{figure}[htp]
    \begin{center}
    \includegraphics[width=8cm]{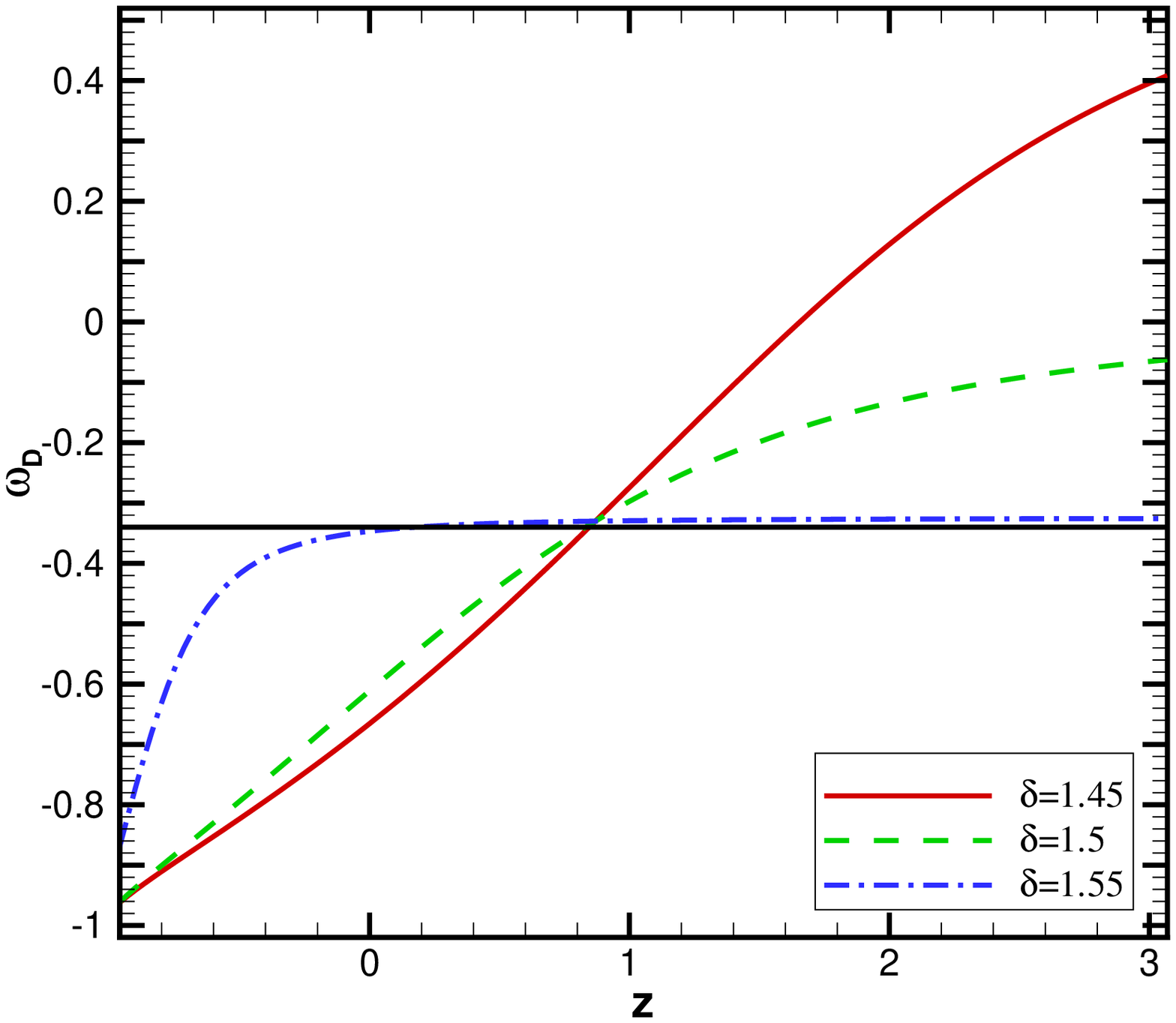}
    \includegraphics[width=8cm]{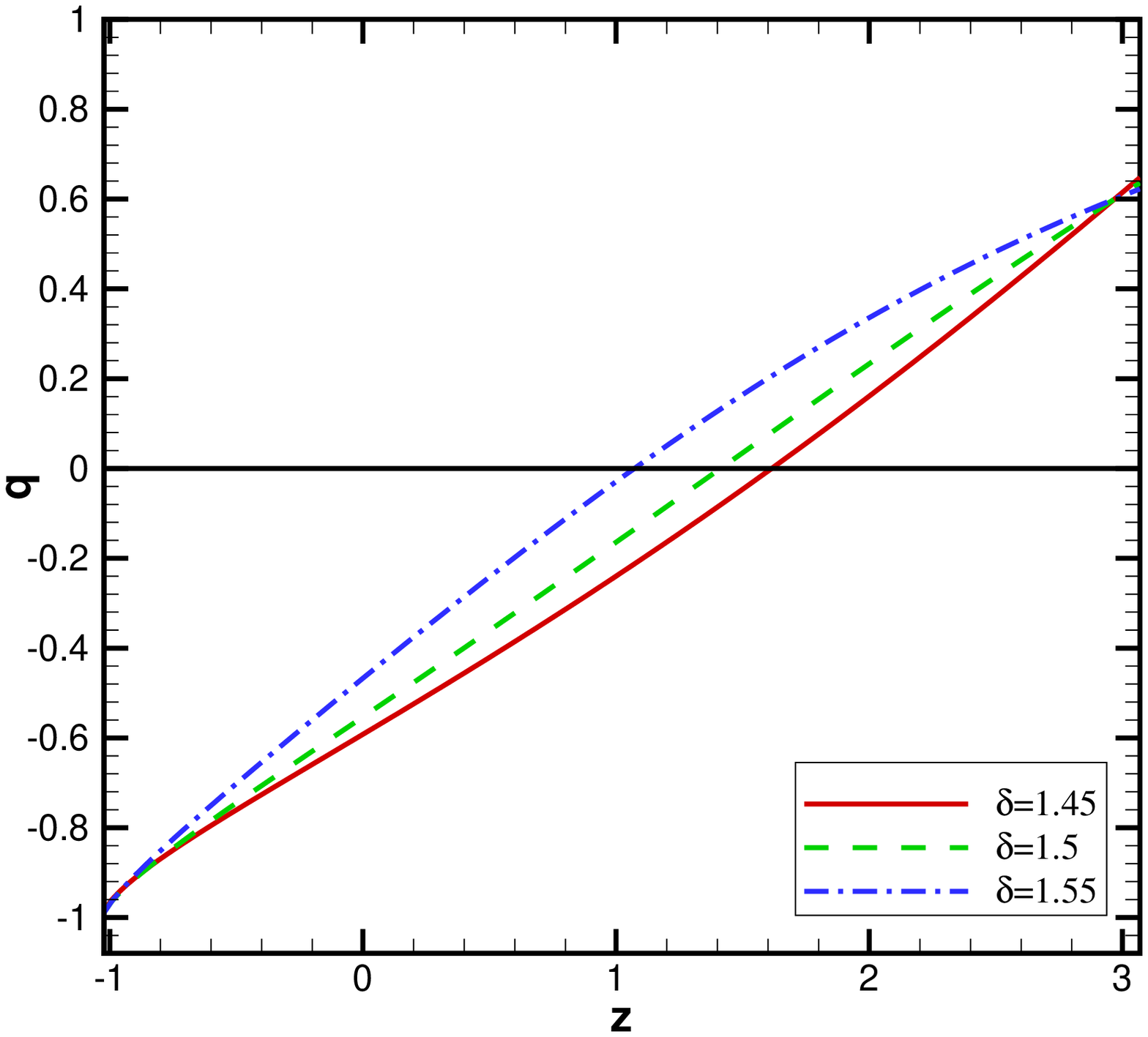}
    \caption{$\omega_D$ and $q$ versus $z$ for $\Omega_{D0}=0\cdot73$ and some values of $\delta$ and $u_0=0\cdot3$.}\label{figw1}
    \end{center}
\end{figure}
In Figs.~\ref{Omega1} and~\ref{figw1}, the behavior of the
dimensionless density, EoS and deceleration parameters have been
plotted against redshift $z$ by considering
$\Omega_{D0}=0\cdot73$ and $u_0=0\cdot3$ for the current
universe. As it is apparent from Fig.~\ref{figw1} and also confirmed
by Eq.~(\ref{w1}), we have $w_D\approx-1$ for DE dominant regime (or equally $u\approx0$) which means that THDE in a cyclic universe
simulates the cosmological constant model of DE at the late time.
The results of employing original holographic dark energy model \cite{HDE5} in cyclic cosmology are also recovered
at the $\delta=1$ limit \cite{Cyclic2}. In summary, the phantom line
is not crossed in this model ($w_D\geq-1$), and the transition
redshift $z_t$ from a deceleration phase to an accelerated universe
lies within the interval $0\cdot35<z_t<1$.

%%%%%%%%%%%%%%%%%%%%%%%%%%%%%%%%%%%%%%%%%%%%%%%%%%%%%%%%%%%%%%%%%%%%%%%%
\section{THDE in DGP braneworld}

For a flat FRW brane embedded in a Minkowski bulk, the Friedmann
equation takes the form \cite{Deffayet,Copeland}

\begin{equation}\label{Friedeq1}
H^2=\Big(\sqrt{\frac{\rho}{3M_{\rm
pl}^2}+\frac{1}{4r^2_c}}+\frac{\epsilon}{2r_c}\Big)^2,
\end{equation}

\noindent where $\rho$ includes the energy density of DM, $\rho_m$,
and DE, $\rho_D$, on the brane, and $r_c=\frac{M_{\rm
pl}^2}{2M_5^3}=\frac{G_5}{2G_4}$ denotes the crossover length scale
between the small and large distances \cite{Deffayet}. It is obvious
that this equation is reduced to
\begin{equation}
H^2=\frac{\rho}{3M_{\rm pl}^2},
\end{equation}
for $r_c\gg1$, nothing but the standard Friedmann equation in flat
FRW spacetime. Eq.~(\ref{Friedeq1}) can also be written as
\begin{equation}\label{Friedeq2}
H^2-\frac{\epsilon}{r_c}H=\frac{\rho}{3M_{\rm pl}^2},
\end{equation}
which reduces to
\begin{equation}\label{RS}
H^2=\frac{\rho^2}{36M_5^6},
\end{equation}
for $\epsilon=-1$ and $r_c \ll H^{-1}$ \cite{RSII}. This result
clearly proves that this branch does not give the self-accelerating
solution which compels us to consider a DE component on the brane to
describe the current accelerated universe. Using Eq.~(\ref{Omega})
and $\Omega_{r_c}=\frac{1}{4H^2r_c^2}$, one can rewrite
Eq.~(\ref{Friedeq2}) as
\begin{eqnarray}\label{Friedeq3}
\Omega_m+\Omega_D+2\epsilon\sqrt{\Omega_{r_c}}=1.
\end{eqnarray}

\begin{figure}[h!]
\begin{center}
\includegraphics[width=8cm]{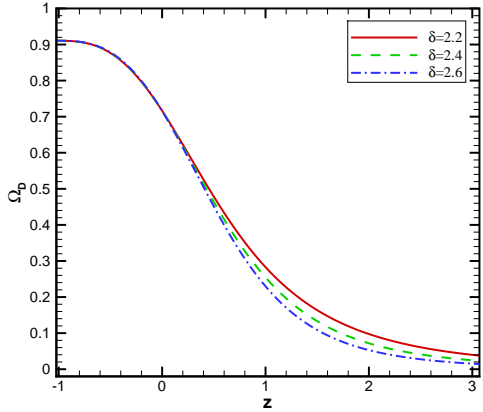}
\caption{The evolution of $\Omega_D$ versus $z$ for $\Omega_{D0} =
0\cdot73 $, $\Omega_{r_c0}=0\cdot002 $ \cite{Xu} and some values of
$ \delta$.}\label{Omega2}
    \end{center}
\end{figure}

For a THDE~(\ref{rhol}) with the Hubble radius as IR cut off
$(L=H^{-1})$, by using~(\ref{Omega}), we obtain
\begin{equation}\label{OmegaD}
\Omega_D=\frac{BH^{2-2\delta}}{3M^2_p}.
\end{equation}
Bearing in mind the time derivative of Eq.~(\ref{rhol})
\begin{equation}\label{rhodot2}
\dot{\rho}_D=2(2-\delta)\rho_D\frac{\dot{H}}{H},
\end{equation}
and combining it with Eq.~(\ref{OmegaD}) and its time
derivative, one finds
\begin{equation}\label{dotOmega2}
\Omega_D^{\prime}=2\Omega_D(1-\delta)\frac{\dot{H}}{H^2},
\end{equation}
where prime denotes the derivative respect to $x=\ln a$, and we used
$\dot{\Omega}_D=H\Omega_D^{\prime}$ to write the above relation.
Now, combining the time derivative of Eq.~(\ref{Friedeq2}) with
Eqs.~(\ref{ConserveCDM}),~(\ref{Omega}) and~(\ref{rhodot2}), we get
\begin{equation}\label{Hdot2}
\frac{\dot{H}}{H^2}=\frac{-3(1-\Omega_D-2\epsilon\sqrt{\Omega_{r_c}})}
{2(\delta-2)\Omega_D-2\epsilon\sqrt{\Omega_{r_c}}+2},
\end{equation}
which can be inserted into~(\ref{dotOmega2}), to reach at
\begin{equation}\label{Omegaprime2}
\Omega_D^{\prime}=\frac{3\Omega_D(1-\delta)(\Omega_D+2\epsilon\sqrt{\Omega_{r_c}}-1)}
{(\delta-2)\Omega_D-\epsilon\sqrt{\Omega_{r_c}}+1}.
\end{equation}
In the limiting case $r_c\gg1$ (or equally $\Omega_{r_c}\rightarrow
0$), $\Omega_D$ of THDE in the Einstein theory \cite{Tavayef} is
restored, a desired result recovering the original HDE ($\Omega_D
=const$) for $\delta=1$. The evolution of $\Omega_D$ as a function
of redshift $z$ is plotted in Fig.~(\ref{Omega2}) for different values
of the parameter $\delta$, whenever $\epsilon=1$,
$\Omega_D(z=0)\equiv\Omega_{D0}=0\cdot73$ and
$\Omega_{r_c}(z=0)=0\cdot002$ \cite{Xu}. Clearly, this figure
indicates that we have $\Omega_D\rightarrow0$ and
$\Omega_D\rightarrow1$, at the early Universe ($z\rightarrow\infty$)
and the late time ($z\rightarrow -1$), respectively.

\begin{figure}[h!]
    \begin{center}
   \includegraphics[width=8cm]{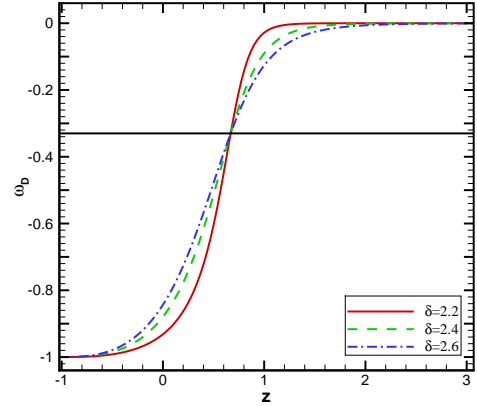}
   \includegraphics[width=8cm]{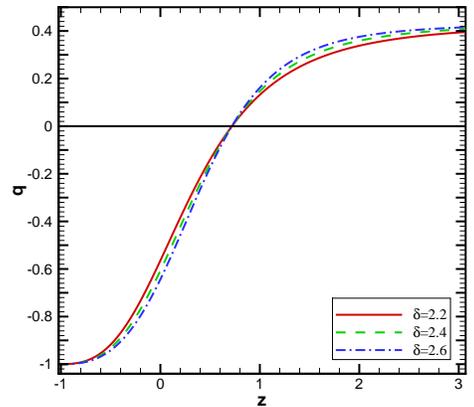}
   \caption{The evolution of the EoS parameter $\omega_D$ and deceletarion parameter $q$ versus
   redshift parameter $z$ for Tsallis HDE in DGP braneworld.
   We have taken $\epsilon=1 $ and  $\Omega_{r_c}=0\cdot002 $\cite{Xu} as the initial conditions. }\label{figw2}
    \end{center}
\end{figure}

Calculations of the EoS and deceleration parameters lead to
\begin{equation}\label{w2}
\omega_D=-1+\frac{(2-\delta)(1-\Omega_D-2\epsilon\sqrt{\Omega_{r_c}})}{(\delta-2)\Omega_D-\epsilon\sqrt{\Omega_{r_c}}+1},
\end{equation}
and
\begin{equation}\label{q2}
q=-1+\frac{3(1-\Omega_D-2\epsilon\sqrt{\Omega_{r_c}})}
{2(\delta-2)\Omega_D-2\epsilon\sqrt{\Omega_{r_c}}+2},
\end{equation}
respectively, which are plotted in Fig.~\ref{figw2}. One can easily see that
for $ r_c\gg 1 $ $ (\Omega_{r_c} \rightarrow 0) $, where the effects
of extra dimension are negligible, the general relativity is
recovered, and hence, Eqs.~(\ref{w2}) and~(\ref{q2}) decrease to
their respective relations \cite{Tavayef}. It is worth
mentioning that, in the limiting case $\delta=1$, the relations of
Ref.~\cite{Ghaffari}, as the desired result, are obtained. From
Fig.~\ref{figw2}, it is obvious that the model can cover the current
accelerated universe even in the absence of a interaction between DM
and DE. We also see that $\omega_D(z\rightarrow-1)\rightarrow-1$ which means
that the model mimics the cosmological constant behavior at future.
The transition redshift ($z_t$) from the acceleration phase to an
accelerated phase lies within the $0\cdot5<z_t<0\cdot9$ range, which
is completely consistent with the recent observations
\cite{Daly,Komatsu,Salvatelli}.
%%%%%%%%%%%%%%%%%%%%%%%%%%%%%%%%%%%%%%%%%%%%%%%%%%%%%%%%%%%%%%%%%%%%%
\section{THDE in RS II braneworld}
\begin{figure}[htp]
\begin{center}
    \includegraphics[width=8cm]{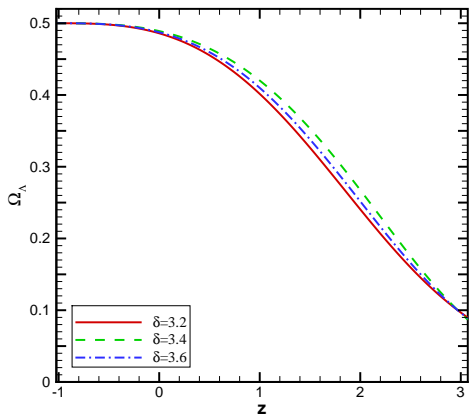}
    \caption{$\Omega_D$ for THDE in RS II braneworld. Here, we have taken $\Omega_{D0}=0\cdot73$
    as the initial condition.}\label{Omega3}
\end{center}
\end{figure}

In RS II braneworld scenario, the modified Friedmann equation on the
brane is written as
\begin{equation}\label{Friedeq6}
H^2+\frac{k}{a^2}=\frac{8\pi}{3M_p^2}\rho+\frac{8\pi}{3M_p^2}\rho_\Lambda
\end{equation}
where $\rho$ denotes the total energy density of the pressureless
source, $\rho_m$, and DE, $\rho_\Lambda$, on the brane, and $
M_p^2=\frac{1}{8\pi G} $ is the reduced Planck mass. Following
\cite{RS}, the energy density of the four dimensional effective DE
is given by
\begin{equation}\label{rho4D}
\rho_\Lambda\equiv\rho_{\Lambda 4}=\frac{M_p^2}{32\pi M_5^3}\rho_{\Lambda 5}+
\frac{3M_p^3}{2\pi\Big(\frac{L_5}{8\pi}-2r_c\Big)^2},
\end{equation}
where $ \rho_{\Lambda 5} $ is the 5D bulk holographic dark energy,
whichfor Tsallis HDE takes the following form
\begin{equation}\label{THDERS}
\rho_{\Lambda 5}=\frac{3c^2B}{4\pi}M_5^3L^{2\delta-4},
\end{equation}
combined with Eq.~(\ref{rho4D}) to get
\begin{equation}\label{rho3}
\rho_\Lambda=\frac{3Bc^2M_p^2}{128\pi^2}L^{2\delta-4}+
\frac{3M_p^3}{2\pi\Big(\frac{L_5}{8\pi}-2r_c\Big)^2},
\end{equation}
for the effective $4$D THDE density. We can eliminate the second
term in relation (\ref{rho3}) for large values of $L$. Moreover,
since $\rho_\Lambda\equiv\rho_{\Lambda 4}$ \cite{RS}, we have
\begin{equation}\label{rho4}
\rho_\Lambda=\frac{3Bc^2M_p^2}{128\pi^2}H^{4-2\delta}.
\end{equation}
Using the definition~(\ref{Omega}), one can write
Eq.~(\ref{Friedeq6}) as follows
\begin{equation}\label{Friedeq7}
1+\Omega_k=\Omega_m+2\Omega_\Lambda,
\end{equation}
where
\begin{equation}\label{OmegaRS}
\Omega_\Lambda=\frac{Bc^2}{16\pi}H^{2-2\delta}.
\end{equation}
Since the WMAP data indicates a flat FRW universe \cite{roos}, we
focus on the $k=0$ case from now on. In this manner,
Eq.~(\ref{Friedeq7}) indicates that whenever $\Omega_m$ is
negligible, $\Omega_\Lambda$ gains its maximum value ($\frac{1}{2}$).
Now, it is a matter of calculations to show that
\begin{equation}\label{Hdot3}
\frac{\dot{H}}{H^2}=\frac{3(2\Omega_\Lambda-1)}{1+2\Omega_\Lambda(\delta-2)},
\end{equation}
and
\begin{equation}\label{Omegaprime3}
\Omega^{\prime}_\Lambda=\Omega_\Lambda(1-\delta)\frac{3(2\Omega_\Lambda-1)}{1+2\Omega_\Lambda(\delta-2)}.
\end{equation}
The behavior of $\Omega_D$ against $z$ is plotted in
Fig.~\ref{Omega3}, where the initial condition
$\Omega_D(z=0)\equiv\Omega_{D0}=0\cdot73$ has been considered. From
this figure we clearly see that at the early universe
($z\rightarrow\infty$) we have $\Omega_D\rightarrow0$, while at the
late time ($z\rightarrow-1$), while DE will dominate and DM is
ignorable, we have $\Omega_D\rightarrow0\cdot5$ in full agreement
with Eq.~(\ref{Friedeq7}).

\begin{figure}[htp]
\begin{center}
    \includegraphics[width=8cm]{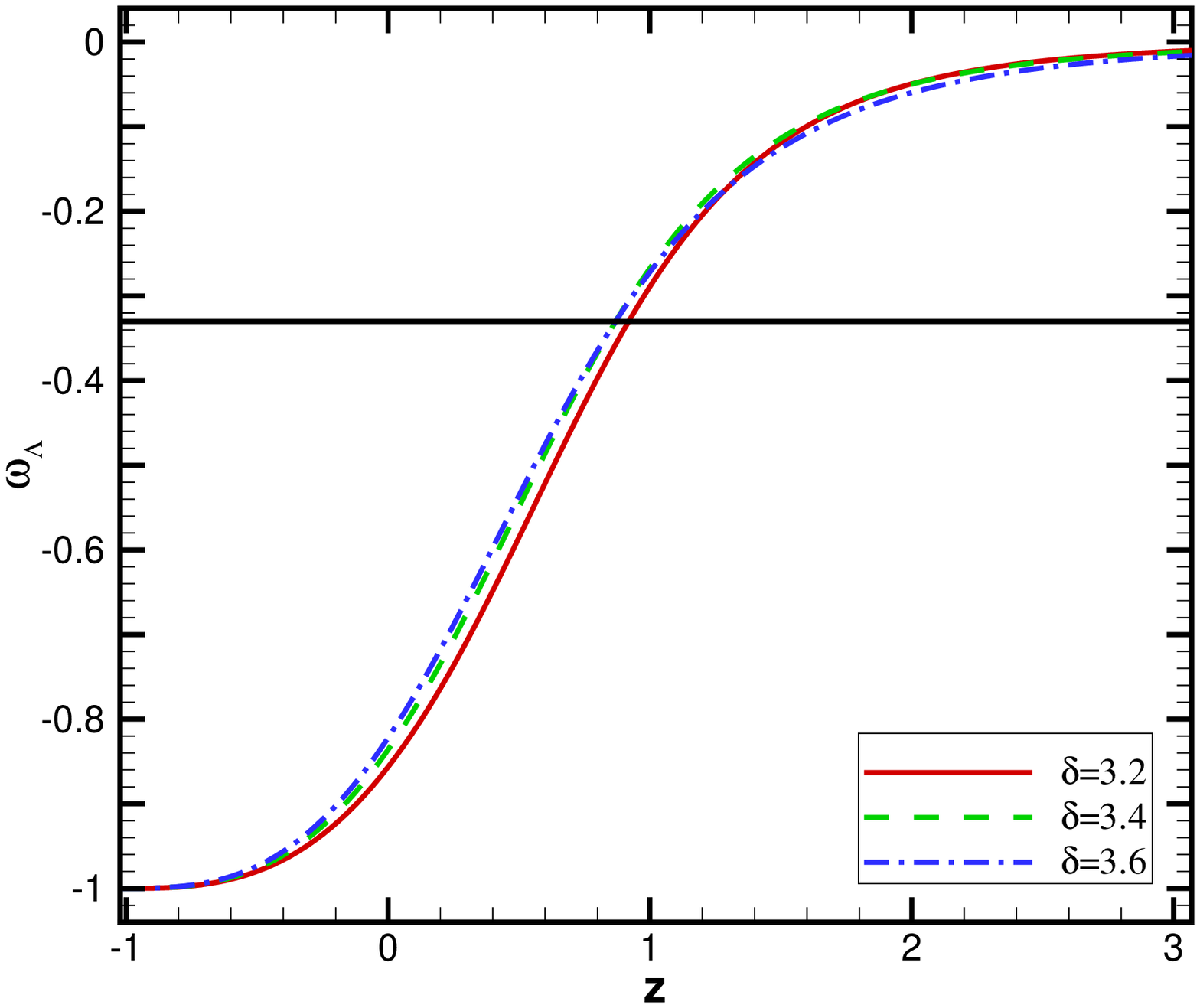}
    \includegraphics[width=8cm]{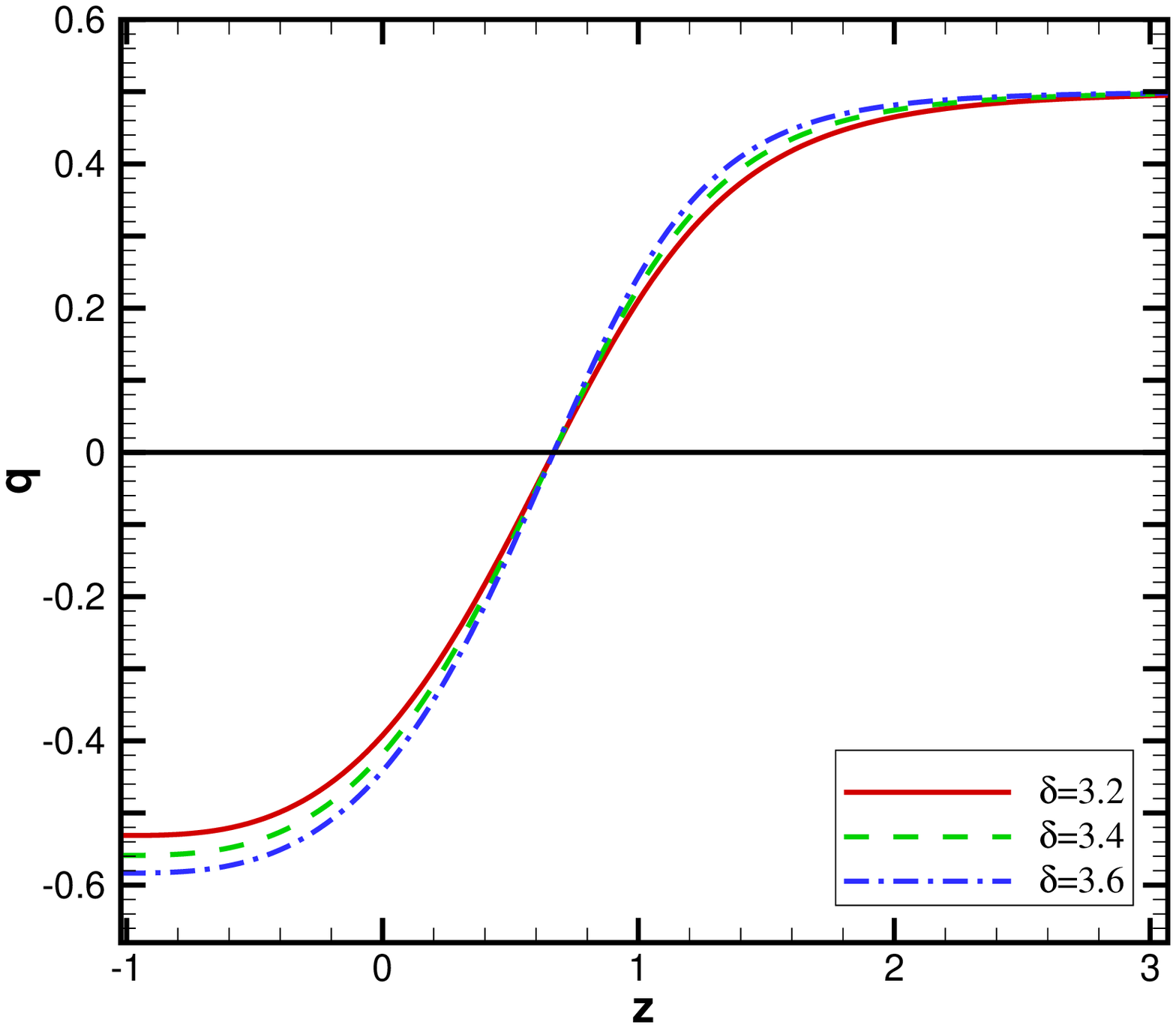}
    \caption{$\omega_D$ and $q$ versus
    $z$ for THDE in RS II brane for some values of $\delta$.}\label{figw3}
\end{center}
\end{figure}

For the EoS and deceleration parameters, one obtains
\begin{equation}\label{w3}
\omega_\Lambda=\frac{1-\delta}{1+2\Omega_\Lambda(\delta-2)},
\end{equation}
and
\begin{equation}\label{q3}
q=-1-\frac{\dot{H}}{H^2}=-1+\frac{3(2\Omega_\Lambda-1)}{1+2\Omega_\Lambda(\delta-2)},
\end{equation}
respectively. They are also depicted for different values of
$\delta$ in Fig.~\ref{figw3}. Our results indicate that the THDE
model with the Hubble cutoff in the RSII braneworld can model the
current accelerated universe, and admits the $0\cdot5<z_t<0\cdot9$
interval for the transition redshift, a result in full accordance
with the recent observations \cite{Daly,Komatsu,Salvatelli}.
\section{Stability}
%%%%%%%%%%%%%%%%%%%%%%%%%%%%%%%%%%%%%%%%%%%%%%%%%%%%%%%%%%%%%%%%%%%%%%%%%%%%%%%%%%%%%%%%%%%%%%%%%%
In this section we would like to study the stability of models
against small perturbations by using the squared of the sound speed
($v_s^2$). In fact, it can be found out by finding the sign of
$v_s^2$. For $v_s^2>0$ the given perturbation propagates in the
environment, and thus, the model is stable against perturbations.
The squared sound speed $v_s^2$ is given by
\begin{equation}\label{v_s}
v_s^2=\frac{dp}{d\rho_D}=\frac{\dot{p}}{\dot{\rho}_D},
\end{equation}
where $P=P_D=\omega_D\rho_{D}$, and finally, we get
\begin{equation}\label{V}
v_s^2=\omega_D+\frac{\dot{\omega}_D\rho_D}{\dot{\rho}_D}.
\end{equation}

%%%%%%%%%%%%%%%%%%%%%%%%%%%%%%%%%%%%%%%%%%%%%%%%%%%%%%%%%%%%%%%%%%%%%%%%%%%%%%%%%%%%%%%%%%%%%%%%%%
\subsection{THDE in Cyclic univrse}
\begin{figure}[htp]
\begin{center}
    \includegraphics[width=8cm]{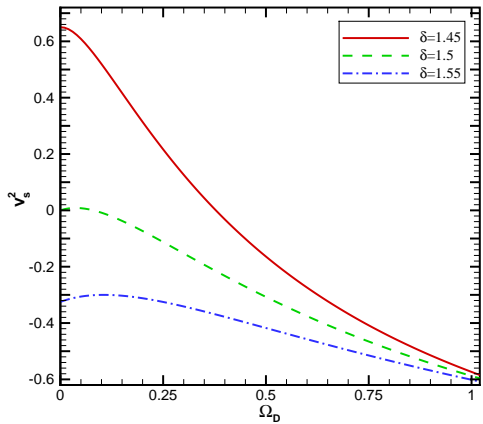}
    \caption{$v_s^2$ versus
    $\Omega_D$ for THDE in cyclic universe. }\label{figV1}
\end{center}
\end{figure}

By taking the time derivative of Eq.~(\ref{w1}) and combining the
result with Eqs.~(\ref{Hdot1}),~(\ref{Omegac2}) and~(\ref{V}), we
can finally obtain the explicit expression of $v_s^2$ for THDE in
cyclic universe. Since this expression is too long, we do not
demonstrate it here, and we only plot it in Fig.~\ref{figV1} showing
that, depending on the values of $\delta$, THDE in cyclic cosmology
can not meet the stability requirement for all values of $\Omega_D$
(or equally $z$).

%%%%%%%%%%%%%%%%%%%%%%%%%%%%%%%%%%%%%%%%%%%%%%%%%%%%%%%%%%%%%%%%%%%%%
\subsection{THDE in DGP braneworld}
\begin{figure}[h!]
    \begin{center}
      \includegraphics[width=8cm]{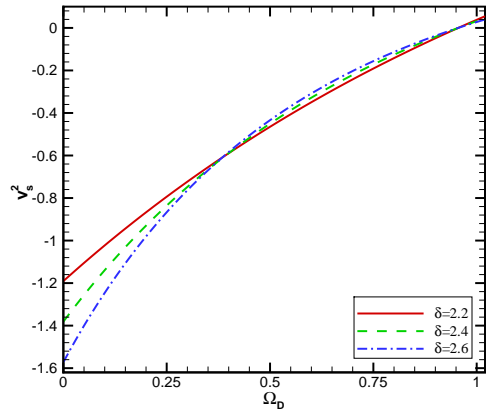}
      \caption{The evolution of $v_s^2$ versus
       $\Omega_D$ for THDE model in DGP braneworld, where
       $\epsilon=+1$ and $\Omega_{r_c0}=0\cdot002 $\cite{Xu}.}\label{figV2}
    \end{center}
\end{figure}

In this manner, calculations lead to
\begin{eqnarray}\label{VDGP}
&&v_s^2=-1+\frac{(2-\delta)(1-\Omega_D-2\epsilon\sqrt{\Omega_{r_c}})}{(\delta-2)\Omega_D-\epsilon\sqrt{\Omega_{r_c}}+1}+\\
&&\frac{\Omega_D(\delta-1)\Big((\delta-2)(1-2\epsilon\sqrt{\Omega_{r_c}})-\epsilon\sqrt{\Omega_{r_c}}+1\Big)}{((\delta-2)\Omega_D-\epsilon\sqrt{\Omega_{r_c}}+1)^2},
\nonumber
\end{eqnarray}
where behavior is shown in Fig~\ref{figV2} against $\Omega_D$. Clearly, we see that
$v_s^2$ is ever negative indicating that THDE in DGP braneworld is
always unstable against the perturbations for $2<\delta$. It is
useful to note here that other values of $\delta$ cannot produce
acceptable behavior for the system parameters including $\Omega_D$,
$q$ and $\omega_D$.

%%%%%%%%%%%%%%%%%%%%%%%%%%%%%%%%%%%%%%%%%%%%%%%%%%%%%%%%%%%%%%%%%%%%%%%%%%%%%%%%
\subsection{THDE in RSII braneworld}
\begin{figure}[h!]
\begin{center}
    \includegraphics[width=8cm]{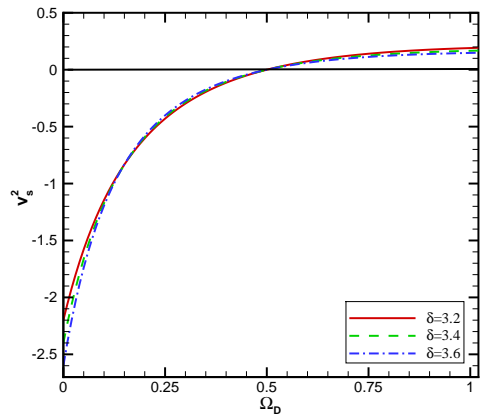}
    \caption{$v_s^2$ versus
    $\Omega_D$ for THDE in RS II braneworld.}\label{figV3}
\end{center}
\end{figure}

Using Eqs.~(\ref{V}),~(\ref{Hdot3}),~(\ref{Omegaprime3}), and the
time derivative of Eq.~(\ref{w3}), one can find
\begin{equation}\label{VRS}
v_s^2=\frac{(1-\delta)(1-2\Omega_D)}{(1+2(2-\delta)\Omega_D)^2},
\end{equation}
where behavior is shown in Fig.~\ref{figV3}. We conclude that THDE in RSII braneworld
is stable for $\frac{1}{2}<\Omega_D<1$ while $\delta>1$. This result is
in agreement with Eq.~(\ref{VRS}), and indeed, this Eq.~(\ref{VRS})
tells that the model is also stable for $0<\Omega_D<1/2$ whenever
$0<\delta<1$.

%%%%%%%%%%%%%%%%%%%%%%%%%%%%%%%%%%%%%%%%%%%%%%%%%%%%%%%%%%%%%%%%%%%
\section{CONCLUSION}
We studied the cosmological consequences of THDE in the Cyclic, DGP
and RS II models. In our study, we focused on a flat FRW brane
filled with a pressureless dark matter and THDE, while there is no
mutual interaction between them. Although all models may describe
the current accelerated universe, none of them are always stable
against small perturbations during the cosmic evolution, at
least at the classical level. In fact, those values of $\delta$
leading the stable models can not produce acceptable behavior for
$\omega_D$, $q$ and $\Omega_D$.
%%%%%%%%%%%%%%%%%%%%%%%%%%%%%%%%%%%%%%%%%%%%%%%%%%%%%%%%%%%%%%%%%%%
\acknowledgments{The work of S. Ghaffari has been supported
financially by Research Institute for Astronomy and Astrophysics of
Maragha (RIAAM). JPMG and IPL thank Coordena\c c\~ao de Aperfei\c
coamento de Pessoal de N\'ivel Superior (CAPES-Brazil), VBB thanks
Conselho Nacional de Desenvolvimento Cient\'ifico e Tecnol\'ogico
(CNPq-Brazil).}
%%%%%%%%%%%%%%%%%%%%%%%%%%%%%%%%%%%%%%%%%%%%%%%%%%%%%%%%%%%%%%%%%%%

\end{document}